\documentclass[aps,prl,twocolumn,superscriptaddress]{revtex4-1}
\usepackage{amsmath}	
\usepackage{graphicx}	
\usepackage{color}
\usepackage{gensymb}

\usepackage{hyperref}
\usepackage{upgreek}

\begin{document}

\title{Optical Valley Hall Effect based on Transitional Metal Dichalcogenide cavity polaritons}

\author{O. Bleu}
\affiliation{Institut Pascal, CNRS/University Clermont Auvergne, 4 avenue Blaise Pascal, 63178 Aubiere, France}

\author{D. D. Solnyshkov}
\affiliation{Institut Pascal, CNRS/University Clermont Auvergne, 4 avenue Blaise Pascal, 63178 Aubiere, France}

\author{G. Malpuech}
\affiliation{Institut Pascal, CNRS/University Clermont Auvergne, 4 avenue Blaise Pascal, 63178 Aubiere, France}

\date{\today}

\begin{abstract}
We calculate the dispersion of spinor exciton-polaritons  in a planar microcavity with its active region containing a single Transitional Metal Dichalcogenide (TMD) monolayer, taking into account excitonic and photonic spin-orbit coupling. We consider the radial propagation of polaritons in presence of disorder. We show that the reduction of the disorder scattering induced by the formation of polariton states allows to observe an optical Valley Hall effect, namely the coherent precession of the locked valley and polarization pseudospins leading to the formation of spatial valley-polarized domains.
\end{abstract}


\maketitle

2D materials represent an enormous emergent field of research in modern Physics \cite{Gupta2015,Gomez2016,Mak2016}. TMD monolayers, due to their chemical structure, exhibit a bandgap at optical frequencies with a strong excitonic resonance \cite{MakPRL2010,Berkelbach2013,Ugeda2014}. The oscillator strength of these excitons is so large that it is possible to observe strong light-matter coupling regime with TMD monolayers  \cite{Dufferwiel2015,Sidler2016} up to room temperature \cite{Liu2015,Flatten2016,Lundt2016}. Because of the honeycomb lattice and the absence of the inversion symmetry, the band structure of TMDs is characterized by 2 valleys with opposite Berry curvature at the corners of the Brillouin zone (usually marked $K$ and $K'$). This leads to particular selection rules for optical emission and absorption: each of the two valleys is definitely associated with its own circular polarization of emitted and absorbed photons, which allows valley pumping and detection using circularly polarized light. Another consequence is a new kind of Hall effect: the Valley Hall effect \cite{Mak2014}. In doped samples, electrons from the two valleys, accelerated by an electric field, undergo opposite lateral drift (anomalous velocity) because of the opposite Berry curvature \cite{Xiao2007,Chang2008}. Scattering and re-acceleration of carriers lead to an overall valley current perpendicular to the main electrical current. 

The spin-orbit coupling (SOC) in these materials is also particularly strong. It suppresses spin relaxation and leads to the coupling of spin and valley degrees of freedom \cite{Mak2012}. This has inspired the development of valleytronics \cite{Beenakker2007}: an analogue of spintronics, where the information is stored in the valley degree of freedom of carriers, which can offer a better protection against relaxation \cite{Jones2013,SuzukiR2014,Zhu2014}. 
Its optical counterpart, the opto-valleytronics, is based on the optical pumping, either resonant or non-resonant\cite{Mak2012, Sallen2012, Kioseoglou2012, Lagarde2014}, with circular or linear polarization, which allows to selectively populate either a single valley or a coherent superposition of valleys.  However, the explicit accounting for the coupling to light leads to a coupling between circularly polarized excitons, which can also be seen as the consequence of the long-range electron-hole Coulomb exchange interaction \cite{Yu2014,Glazov2014}. The two spins, and therefore valleys, form a doublet of excitons coupled to linearly polarized TE and TM light modes. In the well-mastered GaAs quantum wells, this polarization splitting scales quadratically versus the wave vector $k$, and, combined with random scattering on disorder, it leads to spin relaxation through the so-called Maialle-Sham mechanism \cite{Maialle1993}, which is analogous to the Dyakonov-Perel spin relaxation mechanism for electrons \cite{Dyakonov2}. In TMD materials, the mechanism is the same, but the splitting scales linearly with $k$, leading to a much faster spin relaxation. Finally, there is no Berry curvature for excitons \cite{Yu2014}, and, being uncharged, they cannot be accelerated by an electric field, which makes difficult to use a mechanism similar to the electronic one to spatially separate the spin/valley polarized excitons, initially prepared as a mixture and to observe the valley Hall effect. 

A solution for spatial separation of excitons with different polarization, first proposed in \cite{Kavokin2005}, consists in the creation of a radial flow of polarized excitons from a localized source (optical pumping spot) giving rise to the coherent precession of their pseudo-spin induced by the wavevector-dependent effective magnetic field associated with the excitonic SOC. To be observable, this coherent precession should be faster than the scattering by disorder and the decay. In practice, this has been achieved by embedding QWs in an optical cavity in the strong coupling regime, which leads to the formation of a lower polariton branch (LPB) characterized by a small effective mass \cite{Microcavities}, high velocity, and considerably reduced disorder scattering. The LPB usually lies well below the electronic states. In GaAs-based samples, the energy splitting between the TE and TM-polarized eigenmodes is also quantitatively enhanced by the coupling to the photonic modes \cite{Panzarini99,Shelykh2010}. As a result, the system is not anymore in the collisional broadening regime but in a regime where the coherent spin precession can be observed along large distances, and the TE-TM splitting is not anymore a source of spin relaxation. This configuration has allowed the description and the observation of the optical spin Hall effect \cite{Kavokin2005,Leyder2007,maragkou2011optical,Lagoudakis2012}, where the coherent precession creates spin domains. It is also at the basis of the  formation of topological spin current in non-simply-connected geometries \cite{Sala2014}, or the proposal for a polariton-based quantum anomalous Hall effect  \cite{Nalitov2014b}. One should note that indirect excitons are also probably able to reach the coherent precession regime \cite{Vishnevsky2013,High2012} thanks to their long lifetime. 
In TMD monolayers, the TE-TM splitting is much larger than in GaAs based samples, but the short exciton lifetime, their small velocity, and substantial inhomogeneous broadening clearly make this regime out of reach for excitons.

In this work, we consider a TMD monolayer embedded in a planar cavity. We derive the peculiar polariton dispersion in this system taking into account the SOC of both excitons and photons, and study their contribution to the total polariton SOC. This dispersion is characterized by the presence of a minimum out of the light cone which opens an important dissipation channel for disorder scattering. We show that this dissipation channel can be suppressed by working at negative exciton-photon detuning which allows to reach the coherent spin precession regime. Our simulations confirm the possibility of optical observation of the separation of spins and therefore valley polarization in real space -- the optical valley Hall effect (OVHE).

\emph{The model.} We describe the strong coupling of excitons and photons in a planar cavity containing a 2D monolayer of TMD using the coupled oscillator model, where the interaction of the circular polarized spin components of excitons and photons is determined by the light-matter coupling constant $V=\hbar\Omega_R/2$, where $\Omega_R$ is the Rabi frequency. We consider the case of weak residual doping which allows to neglect the the trion resonance.
The SOC for photons arises from the cavity TE-TM splitting and is quadratic in $k$ and with a double winding in $\varphi$. Qualitatively, it means that transverse polarization is the same for two opposite propagation directions. The SOC for excitons in TMDs is defined by the symmetry of the lattice and by the optical selection rules, which give a double winding in the polar angle $\varphi$, but with a splitting linear in $k$ \cite{Yu2014,Glazov2014}. However one should note that Ref. \cite{Yu2014} predicts $k$-linear splitting in the whole reciprocal space, whereas Ref. \cite{Glazov2014} predicts it only beyond the light cone. We use the description of Ref.\cite{Yu2014},  keeping in mind that the scheme we propose can allow to decide between the two results. The characteristic strength of the SOC is described by the constants $\alpha$ (for excitons) and $\beta$ (for photons). The Hamiltonian of the coupled exciton-photon system (see \cite{suppl} for the discussion of the full Hamiltonian) in the circular polarization basis reads:
\begin{equation}
\hat H = \left( {\begin{array}{*{20}{c}}
{{E_{X}(k)}}&{\alpha k{e^{ - 2i\varphi}}}&V&0\\
{\alpha k{e^{ 2i\varphi}}}&{{E_{X}(k)}}&0&V\\
V&0&{{E_{P}(k)}}&{\beta {k^2}{e^{ - 2i\varphi }}}\\
0&V&{\beta {k^2}{e^{2i\varphi }}}&{{E_{P}(k)}}
\end{array}} \right)
\label{Ham}
\end{equation}
where $E_{X}(k)$ and $E_{P}(k)$ are the bare exciton and cavity photon dispersions, which are assumed to be parabolic, with excitonic and photonic masses $m_X$ and $m_P$ respectively.
Diagonalizing this Hamiltonian, we obtain the following dispersion of the four exciton-polariton states:

\begin{eqnarray}
E &=& \frac{1}{2}(E_P(k)+E_X(k)  \pm k\left( \alpha  + \beta k \right)\\ 
&\pm& \sqrt {{k^2}{{\left(E_P(k)-E_X(k)\mp\left( {\alpha  - \beta k}\right) \right)}^2} + 4{V^2}}  )\nonumber
\end{eqnarray}

\begin{figure}[tbp]
\includegraphics[scale=0.33]{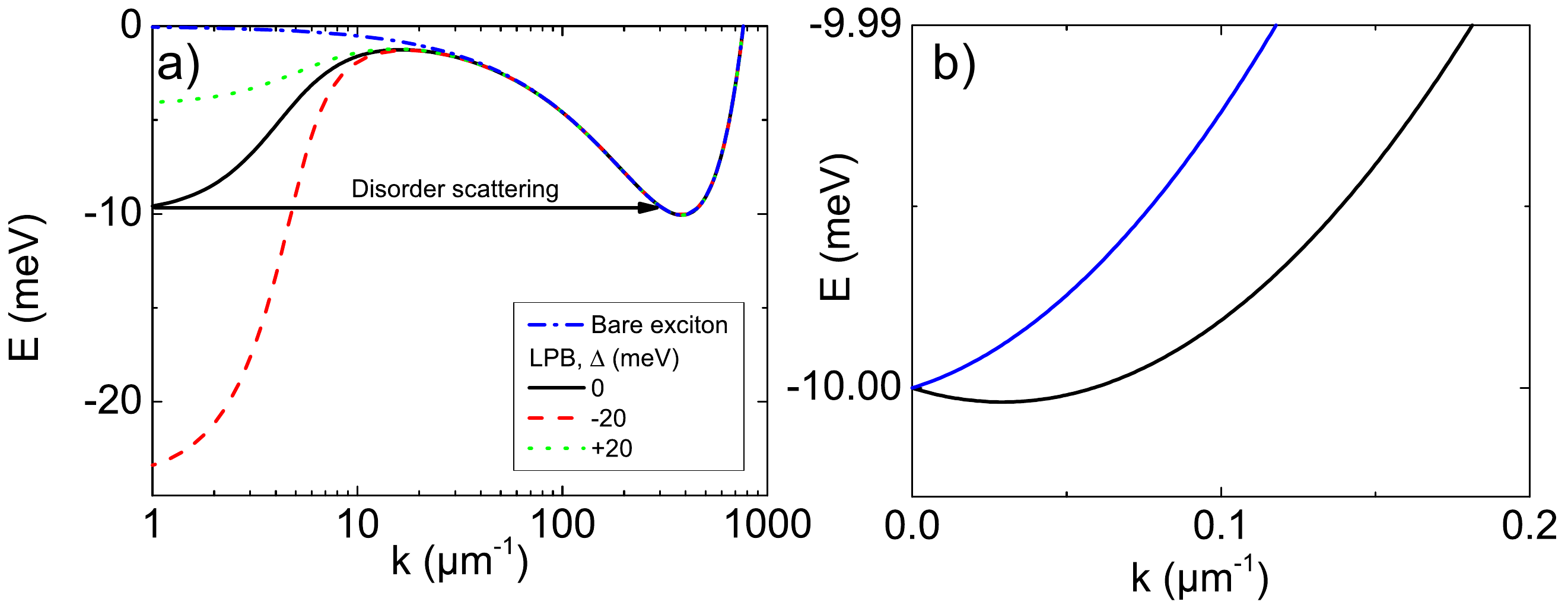}
\caption{\label{figdisp} (Color online) a) Dispersion of lowest energy state of polaritons in TMD for 3 different detunings (solid black, dashed red, and dotted green lines) and the lowest bare exciton branch (dash-dotted blue line). b) Zoom on the minimum of the LPB at low $k$.}
\end{figure}

The dispersion of the lowest energy state of the quadruplet is shown on the Fig. \ref{figdisp}(a) for 3 different values of the detuning (defined as $\Delta=E_{P}(0)-E_{X}(0)$). We have used the following set of parameters throughout the paper: $m_P=4\times 10^{-5} m_0$, $m_X=0.6m_0$ ($m_0=9.1\times 10^{-31}$ kg is the free electron mass), $\alpha=52.6~\mu$eV$/ \mu$m$^{-1}$, $\beta=47.3~\mu$eV$/ \mu$m$^{-2}$, $\hbar\Omega_R=20$ meV. There are several important features in these dispersions.

Because of the $k$-linear dependence of the SOC, the dispersion is linear for bare excitons at low $k$. Since the exciton mass is four orders of magnitude larger than the cavity photon mass, this linear region is quite wide, as compared to the light cone. The lower branch of the exciton dispersion (blue dash-dotted line) exhibits a minimum at $k_{m}=\alpha m_{X}/\hbar^2$, with its energy given by $E_{m}=-\alpha^2 m_{X}/2\hbar^2$. For the typical TMD parameters, this minimum lies out of the light cone, and therefore its position is not affected by the light-matter coupling and by the detuning, as can be seen on Fig. \ref{figdisp}(a), where this minimum appears at around $k=400~\mu$m$^{-1}$ for all detunings. One should notice that this minimum is not affected by the discripancies between \cite{Yu2014} and \cite{Glazov2014}. The energy $E_{m}$ can be expected to be several tens of meV below the bare exciton $E_X(k=0)$, depending on the SOC magnitude in a particular material. As one can see, for a particular range of detunings this minimum is resonant with the strongly coupled polariton states.  Polaritons are typically well-protected from the excitonic disorder because of its small characteristic scale $l_X\approx 10 nm$, and because are polaritons are typically blue detuned by half of the Rabi splitting with respect to the bare excitons states. The presence of resonant excitonic states weakens this protection, because such disorder can lead to the scattering of polaritons into high-$k$ states (arrow in Fig. \ref{figdisp}(a)), which corresponds to additional effective losses for polaritons, as we will study in detail in the last part of the manuscript.

Another feature of the dispersion is due to the fact that a linear splitting dominates all other terms at low wavevectors. Therefore, the linear excitonic dispersion is inherited by polaritons (in spite of their light mass) near $k=0$ (within the light cone), as shown on the Fig. \ref{figdisp}(b). One should notice that this minimum does not exist for  a quadratic in  $k$ excitonic SOC \cite{Glazov2014}.  In any case, the depth of the related minimum is expected to be in the $\mu$eV range, which should render its experimental observation difficult. For this reason, we will study only the effects arising from the excitonic minimum present at large $k$.

From the structure of the Hamiltonian we see that the SOCs of both excitonic and photonic origins add up together for the strongly coupled polariton branches, which enhances the resulting TE-TM splitting and favors the observation of the OVHE.

\begin{figure}[tbp]
\includegraphics[scale=0.32]{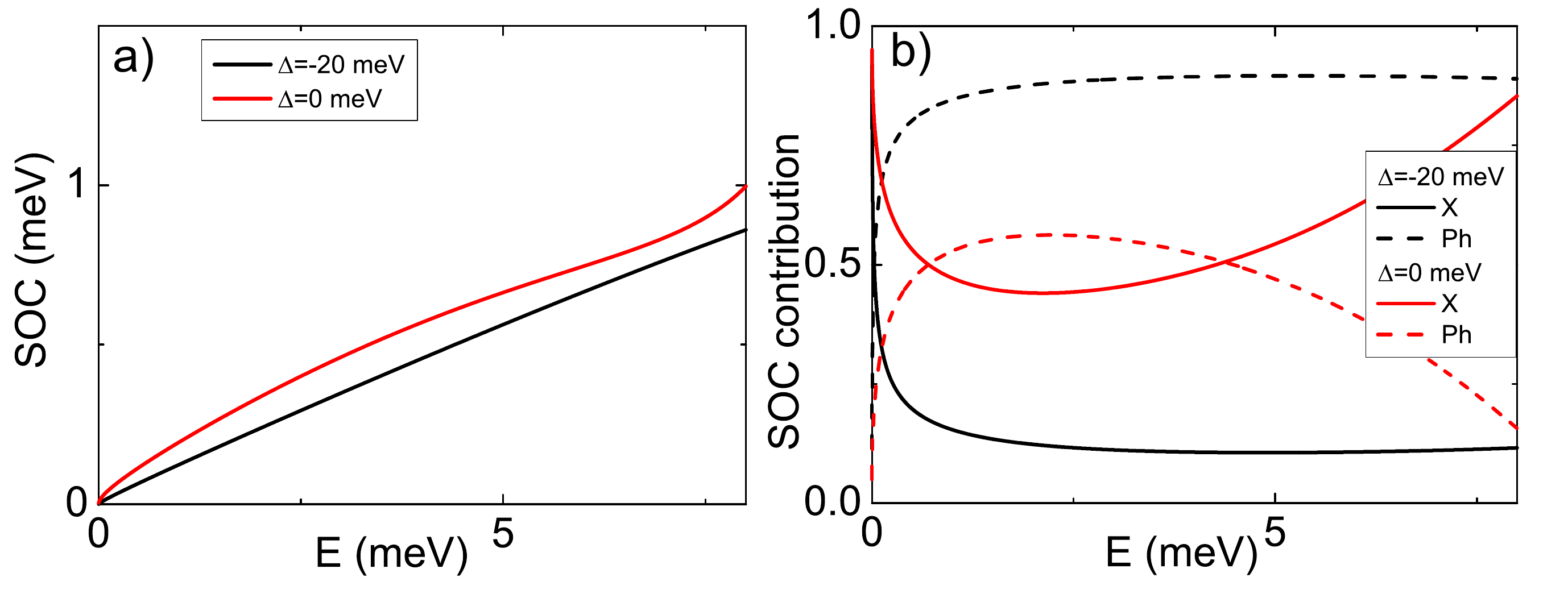}
\caption{\label{figsoc} (Color online) a) The polariton SOC as a function of LPB energy at two detunings; b) The exciton (X) and photon (Ph) contributions to the SOC as a function of LPB energy for the same detunings.}
\end{figure}

The magnitude of the SOC on the polariton lower branch for two different detunings is shown on the Fig. \ref{figsoc}(a).  While at strongly negative detunings (black curve) this dependence is almost linear, at less negative or zero (red curve) detuning it becomes nonlinear, exhibiting an inflection point. This occurs because of the interplay of the excitonic and photonic SOC, which not only depend differently on the wavevector, and therefore, on the LPB energy, but also contribute differently at different energies, according to the varying photonic and excitonic fractions of polariton. These contributions are shown in Fig. \ref{figsoc}(b) for the same detunings as in Fig. \ref{figsoc}(a). At low energies and wavevectors, the excitonic SOC (linear in $k$) always dominates, but at higher energies the result depends on the detuning: at $\Delta=-20$ meV, the photonic SOC quickly becomes dominant and remains such, while at $\Delta=0$ meV, the photonic SOC is comparable with the excitonic one only in a very narrow region, whereas the excitonic SOC dominates everywhere else. This last regime allows to use the polaritonic OVHE to measure the excitonic SOC almost directly, while keeping the benefit of the smaller polariton mass, providing fast propagation.

\emph{Optical Valley Hall Effect}  The optical spin Hall effect \cite{Kavokin2005,Leyder2007,maragkou2011optical,Lagoudakis2012} is based on the coherent precession of the polarization of light during its radial propagation from the injection spot about the $k$-dependent effective magnetic field describing the SOC. For TMD monolayers, the circular polarization has a one-to-one correspondence with the valley degree of freedom, and therefore the circular polarization degree allows a direct insight into the valley polarization, while linear polarization is a signature of a coherent superposition of the populations of the valleys. We therefore consider a resonant light injection in a tight spot with a chosen polarization, for example, linear, in order to excite a superposition of both valleys at a given energy determined by the laser frequency, and then monitor the spatial distribution of the circular polarization degree generated by the radial propagation. 

To demonstrate the possibility of the observation of OVHE with polaritons in TMD, we have carried out numerical simulations, solving coupled Schrodinger equations for excitons and photons, taking into account both spin components, and both types of SOC:

\begin{eqnarray}
i\hbar \frac{{\partial \psi }}{{\partial t}} = \hat T_{ph}\psi  + \hat S_{ph}\psi  - \frac{{i\hbar }}{{2\tau_{ph} }}\psi  + V\phi  + \hat P\\
i\hbar \frac{{\partial \phi }}{{\partial t}} = {\hat T_X}\phi  + {\hat S_X}\phi  - \frac{{i\hbar }}{{2{\tau _X}}}\phi  + U\phi  + V\psi \nonumber
\label{schro}
\end{eqnarray}

Here, $\psi$ is the photon spinor wavefunction, $\phi$ is the exciton spinor wavefunction, $V$ is light-matter coupling, $U$ is the disorder potential acting on the excitons with a correlation length $l_{X}=4$ nm and amplitude 5 meV. $\hat T_{ph,X}$ are the kinetic energy operators for photons and excitons (with the masses $m_{ph,X}$), $\hat S_{ph,X}$ are the SOC operators (with $k^2$ and $k$ dependences and with different coupling constants $\beta$ and $\alpha$, respectively). These operators are written in the reciprocal space \cite{suppl}. $\hat P$ is the $cw$ pumping operator, corresponding to a Gaussian pumping spot and a quasi-resonant frequency. 
The high-resolution simulations, required to describe simultaneously low wavevectors ($10^6$ m$^{-1}$), where the OVHE is taking place, and high wavevectors ($10^9$ m$^{-1}$), where the excitonic minimum is located, and to take into account the small correlation length of the excitonic disorder, were carried out using the nVidia CUDA framework.

\begin{figure}[tbp]
\includegraphics[scale=0.33]{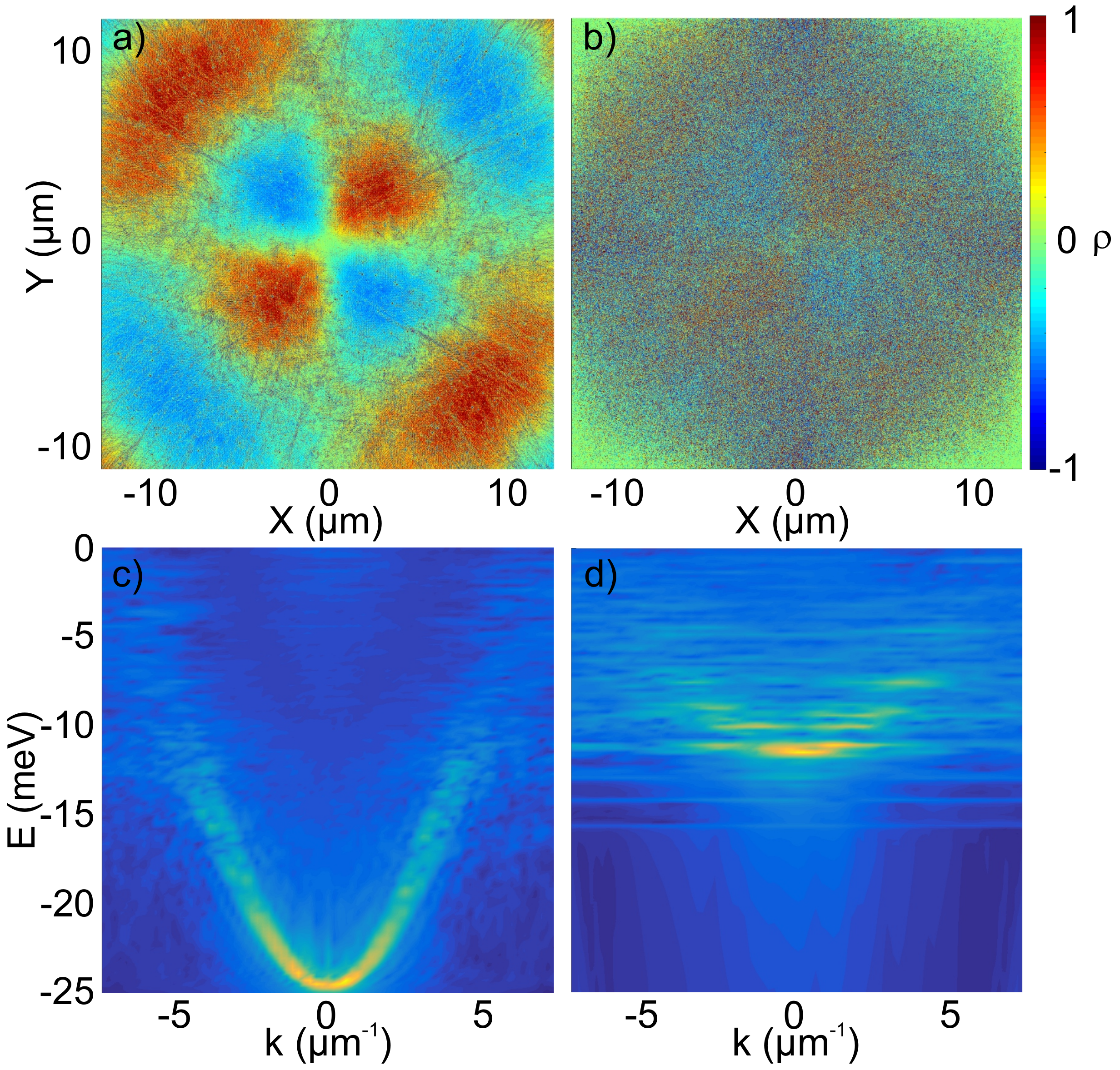}
\caption{\label{figvshe} (Color online) a,b) Spatial image of the circular (valley) polarization degree. c,d) Reciprocal space polariton emission. a,c) negative detuning, optimal for OVHE; b,d) positive detuning, OVHE not observable. }
\end{figure}

The calculated real space propagation images and reciprocal space dispersions are shown in Fig. \ref{figvshe}. Panels (a,c) were obtained at negative detuning $\delta=-20$ meV. The real space image is obtained with the pumping laser detuned 6 meV above the LPB bottom. At this detuning, we see on the dispersion a moderate effect of the disorder between -25 and -12 meV, whereas at higher energies the dispersion strongly broadens because of disorder scattering. Two dispersion lines corresponding to the two polarizations can be resolved. Indeed, the real space simulation demonstrates four clear polarization domains, corresponding to the double winding of the TE-TM splitting, and one full period of rotation. The domain size is of the order of 2 $\mu$m. It is visible thanks to the favorable combination of the polariton parameters with respect to excitons, namely a fast propagation velocity (1 $\mu$m/ps) and a reduced spin relaxation time (4 ps). Panels (b,d), are calculated at $\delta=0$ meV, when the LPB is resonant with the exciton reservoir. The polariton dispersion can still be observed, but it becomes very broad. The real space propagation is essentially dominated by noise. The polarisation degree  is below 5\%. Therefore, the observation of the OVHE in TMD cavities could be possible at sufficiently negative detunings.

\begin{figure}[tbp]
\includegraphics[scale=0.3]{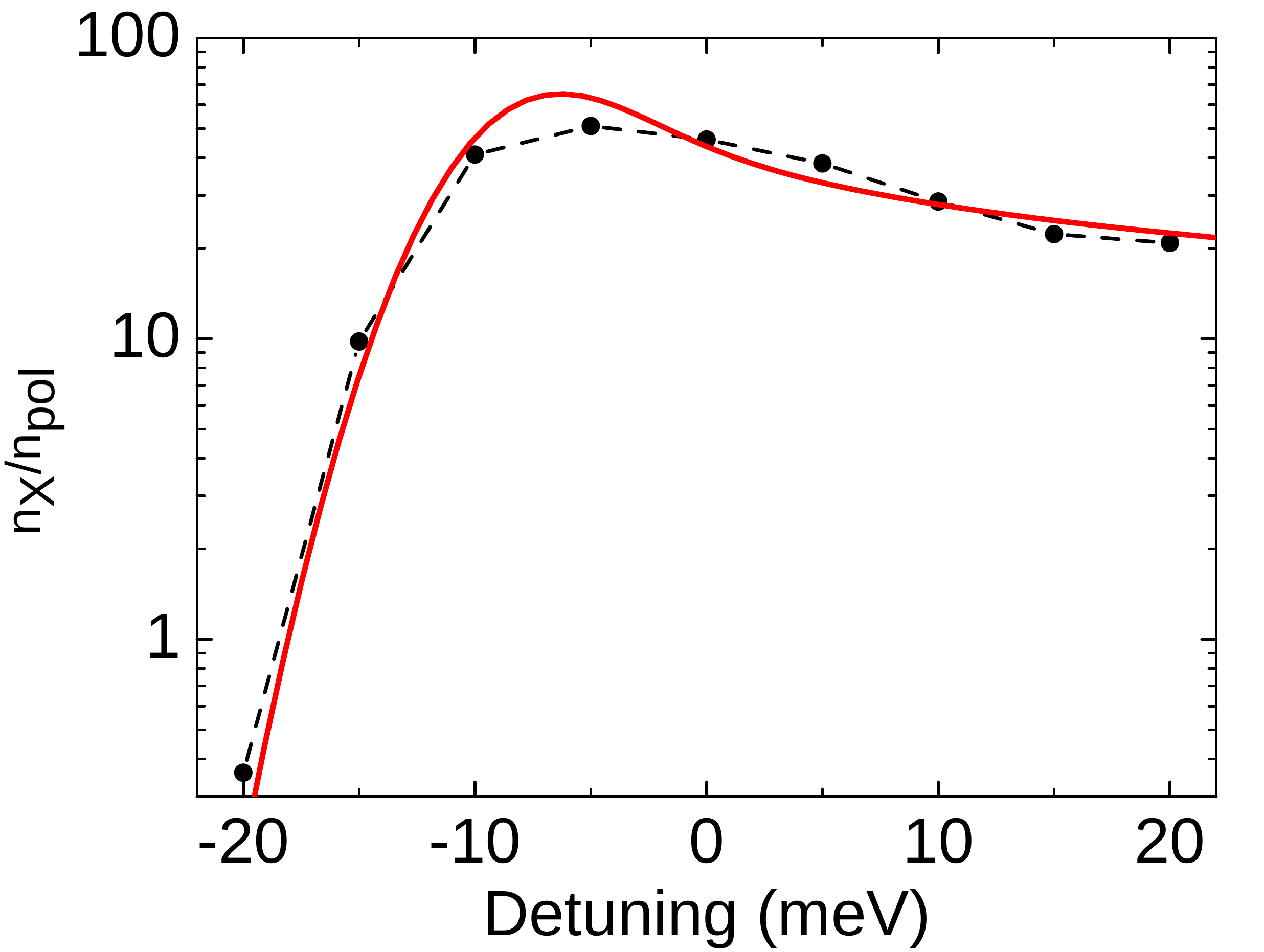}
\caption{\label{figresonance} (Color online) The ratio between scattered $n_X$ and remaining density $n_{pol}$ as a function of detuning: black dots - numerical simulations, red solid line - analytical DOS.}
\end{figure}

\emph{Disorder scattering}
In the last part we quantitatively analyse the magnitude of losses induced by the substantial disorder scattering present in real samples \cite{Dufferwiel2015}.
Figure \ref{figresonance}  shows the density transfer from polaritons to excitonic states versus detuning. The black points were obtained from the solution of the equation \eqref{schro}. Polariton states were excited with a \emph{CW} pump with $\sigma=0.4~\mu$m, set 2 meV above the bottom of the LPB, in order to obtain spatial propagation required for OVHE \cite{suppl}. The simulation is run for several hundreds of ps, until the particle distribution is settled. A clear maximum is observed when the bottom of the polariton dispersion is resonant with the excitonic minimum at large wave vector.

The scattering rate due to disorder given by the Fermi's golden rule is proportional to the final density of states (DOS)
\begin{equation}
\Gamma_{pol\to X}=\frac{2\pi}{\hbar} |V_k|^2 \rho(E),
\end{equation}
where $k$ is the wavevector of a high-$k$ exciton state resonant with the bottom of the polariton dispersion, $V_k$ is the matrix element of the disorder scattering, $|V_k|^2\propto \exp(-k^2l_X^2)$, and $\rho(E)$ is the DOS given by:
\begin{equation}
\rho\left(E\right)=\frac{2\pi m}{\hbar^2}\left(1+\frac{\alpha}{\sqrt{\alpha^2+2E\hbar^2/m}}\right)
\label{eqrho}
\end{equation}
It diverges at $E=E_m$ because of the excitonic minimum at nonzero $k$, which explains a strong scattering resonance at this energy. Taking into account the Gaussian inhomogeneous broadening $\sigma$ by convolution with $\rho(E)$ removes the divergency and gives the following DOS:
\begin{equation}
\rho \left( E' \right) = \frac{{\sqrt{\pi \left| E' \right|} }}{{2\sigma \sqrt 2 }}{e^{ - \frac{{{E'^2}}}{{2{\sigma ^2}}}}} \left( {{I_{ - \frac{1}{4}}}\left( {\frac{{{E'^2}}}{{2{\sigma ^2}}}} \right) + \frac{E'}{|E'|} {I_{\frac{1}{4}}}\left( {\frac{{{E'^2}}}{{2{\sigma ^2}}}} \right)} \right)
\end{equation}
where $I$ are the modified Bessel functions of the second kind, and $E'=E-E_m$. The corresponding analytical curve is shown in Fig. \ref{figresonance} in red. It exhibits an asymmetric behavior, because the left tail (negative detuning) exists only due to broadening, and therefore decays as a Gaussian. The right tail of the curve (positive detuning) decays smoothly as $1/\sqrt E$ as the pure DOS.

The scattering to high-$k$ states after resonant optical excitation can be considered as a substantial contribution to the particle decay, shortening their lifetime, because the probability of backward scattering is very low. Our analytical and numerical results demonstrate that while at positive detunings the OVHE might be destroyed by the scattering on the excitonic disorder, it should be possible to observe OVHE in TMD monolayers at sufficiently negative detunings, and still have an important contribution of the exciton SOC to the total polariton SOC. 

To conclude, strong coupling of 2D TMD monolayers in planar cavities allows to observe optical Valley Hall effect. However, due to a peculiar dispersion of excitons and exciton-polaritons arising from the SOC, special care should be taken to avoid resonant scattering to high-$k$ exciton states by disorder.

We thank A.I. Tartakovskii, D.N. Krizhanovskii, and S. Dufferwiel for useful discussions. We acknowledge the support of the ANR QFL.

\section{Supplemental Material}
In this supplemental material we discuss the validity of the reduced Hamiltonian of the main text, deriving it from the full Hamiltonian containing the split-off dark exciton states. We also present the details of the numerical realization of the spin-orbit coupling.

\subsection{Full Hamiltonian}
The full Hamiltonian describing the strong coupling of excitons and photons in Transitional Metal Dichalcogenides (TMDs) includes the two spin projections of the two excitons formed from $K$ and $K'$ valleys ($K^\pm$ and $K'^\pm$) and the two spin projections for photons ($\sigma^+$ and $\sigma^-$). The spin-orbit coupling (SOC) for excitons splits the circular-polarized states by $\Delta_Z\approx 250$ meV. The exchange interaction gives rise to $k$-linear terms $\alpha k e^{2i\varphi}$ coupling the two valleys, but only for the two spin components interacting with light ($K^+$ and $K'^-$). The same two components are strongly coupled by $V$ ($2V=\hbar\Omega_R$, where $\Omega_R$ is the Rabi frequency) with the photons of the same circular polarization. Finally, the two circular photons are coupled by the TE-TM splitting $\beta k^2e^{2i\varphi}$, quadratic in wavevector. On the basis $(X_{K+},X_{K-},X_{K'+},X_{K'-},\Gamma_{+},\Gamma_{-})$, this Hamiltonian reads:

\begin{widetext}
\begin{equation}
\left( {\begin{array}{*{20}{c}}
{{E_X}\left( k \right)}&0&0&{\alpha k{e^{ - 2i\varphi }}}&V&0\\
0&{{\Delta _z} + {E_X}\left( k \right)}&0&0&0&0\\
0&0&{{\Delta _z} + {E_X}\left( k \right)}&0&0&0\\
{ak{e^{2i\varphi }}}&0&0&{{E_X}\left( k \right)}&0&V\\
V&0&0&0&{\Delta  + {E_P}\left( k \right)}&{\beta {k^2}{e^{ - 2i\varphi }}}\\
0&0&0&V&{\beta {k^2}{e^{2i\varphi }}}&{\Delta  + {E_P}\left( k \right)}
\end{array}} \right)
\end{equation}
\end{widetext}

The analysis of the eigenstates of this Hamiltonian shows that the two dark exciton states split off by $\Delta_Z$ do not participate in the formation of the main quadruplet of polariton states, and the properties of the latter are well described by the reduced Hamiltonian written in the main text (Eq. (1)), which contains only bright states.

\subsection{Numerical simulations}
The full Schrodinger equation written in the main text (Eq. (3)) combines terms which can be optimally solved in real space and in reciprocal space:
\begin{eqnarray}
i\hbar \frac{{\partial \psi }}{{\partial t}} = \hat T_{ph}\psi  + \hat S_{ph}\psi  - \frac{{i\hbar }}{{2\tau_{ph} }}\psi  + V\phi  + \hat P\\
i\hbar \frac{{\partial \phi }}{{\partial t}} = {\hat T_X}\phi  + {\hat S_X}\phi  - \frac{{i\hbar }}{{2{\tau _X}}}\phi  + U\phi  + V\psi \nonumber
\label{schro}
\end{eqnarray}
Especially, the particular case of $k$-linear SOC with a double winding is somewhat special, and does not allow a straightforward representation with real-space operators. Indeed, the TE-TM double-winding SOC is usually written in real space as a term proportional to $(\partial/\partial x\pm i\partial/\partial y)^2$, where the derivatives naturally give both the amplitude and the texture of the SOC. This is impossible for double-winding combined with linear $k$. In our case, we use the reciprocal space representation for the kinetic energy and the SOC operators $\hat T$, $\hat S_{ph}$, $\hat S_{X}$:
\begin{eqnarray}
\hat T\psi  &=& {F^{ - 1}}\left( {\frac{{{\hbar ^2}{k^2}}}{{2m}}F\left( \psi  \right)} \right)\\
{\hat S_{ph}}\psi  &=& {F^{ - 1}}\left( {\beta {{\left( {{k_x} \mp i{k_y}} \right)}^2}F\left( \psi  \right)} \right)\\
{\hat S_X}\psi  &=& {F^{ - 1}}\left( {\frac{{\alpha {{\left( {{k_x} \mp i{k_y}} \right)}^2}}}{k}F\left( \psi  \right)} \right)
\end{eqnarray}
where $F$ is the 2D Fourier transform and $F^{-1}$ is the inverse 2D Fourier transform.
Moreover, since the problem includes very high wavevectors, additional approximations are necessary both for photon dispersion and for photon SOC, since the corresponding frequencies become too large, and the parabolic dispersion becomes anyway not valid for photons. The corresponding photonic states are not coupled to the exciton at high $k$, and therefore do not contribute to the physics of the problem. Therefore, we introduce a saturation of the photonic dispersion and of the photonic SOC at a frequency of $10^{15}$ Hz ($\approx 626$ meV above the cavity mode), to make the numerical simulations feasible with a time step of $10^{-16}$ s. The spatial grid size was $2048^2$ points.

\subsection{The role of laser detuning}

An important role is played by the $cw$ laser detuning with respect to the bottom of the polariton branch. Indeed, higher laser detuning means higher kinetic energy, faster propagation time, and less sensitivity to the disorder (smaller $V_k$). Therefore, we can conclude that in order to optimize the observation of OVHE it is better to work at more negative exciton-photon detuning, to avoid being in resonance with high-$k$ states, and at the same time use higher laser detuning, to increase the TE-TM splitting and the propagation speed, while at the same time reducing the effect of the disorder. 

\bibliography{valley_pol,reference}

\begin{thebibliography}{38}%
\makeatletter
\providecommand \@ifxundefined [1]{%
 \@ifx{#1\undefined}
}%
\providecommand \@ifnum [1]{%
 \ifnum #1\expandafter \@firstoftwo
 \else \expandafter \@secondoftwo
 \fi
}%
\providecommand \@ifx [1]{%
 \ifx #1\expandafter \@firstoftwo
 \else \expandafter \@secondoftwo
 \fi
}%
\providecommand \natexlab [1]{#1}%
\providecommand \enquote  [1]{``#1''}%
\providecommand \bibnamefont  [1]{#1}%
\providecommand \bibfnamefont [1]{#1}%
\providecommand \citenamefont [1]{#1}%
\providecommand \href@noop [0]{\@secondoftwo}%
\providecommand \href [0]{\begingroup \@sanitize@url \@href}%
\providecommand \@href[1]{\@@startlink{#1}\@@href}%
\providecommand \@@href[1]{\endgroup#1\@@endlink}%
\providecommand \@sanitize@url [0]{\catcode `\\12\catcode `\$12\catcode
  `\&12\catcode `\#12\catcode `\^12\catcode `\_12\catcode `\%12\relax}%
\providecommand \@@startlink[1]{}%
\providecommand \@@endlink[0]{}%
\providecommand \url  [0]{\begingroup\@sanitize@url \@url }%
\providecommand \@url [1]{\endgroup\@href {#1}{\urlprefix }}%
\providecommand \urlprefix  [0]{URL }%
\providecommand \Eprint [0]{\href }%
\providecommand \doibase [0]{http://dx.doi.org/}%
\providecommand \selectlanguage [0]{\@gobble}%
\providecommand \bibinfo  [0]{\@secondoftwo}%
\providecommand \bibfield  [0]{\@secondoftwo}%
\providecommand \translation [1]{[#1]}%
\providecommand \BibitemOpen [0]{}%
\providecommand \bibitemStop [0]{}%
\providecommand \bibitemNoStop [0]{.\EOS\space}%
\providecommand \EOS [0]{\spacefactor3000\relax}%
\providecommand \BibitemShut  [1]{\csname bibitem#1\endcsname}%
\let\auto@bib@innerbib\@empty
\bibitem [{\citenamefont {Gupta}\ \emph {et~al.}(2015)\citenamefont {Gupta},
  \citenamefont {Sakthivel},\ and\ \citenamefont {Seal}}]{Gupta2015}%
  \BibitemOpen
  \bibfield  {author} {\bibinfo {author} {\bibfnamefont {A.}~\bibnamefont
  {Gupta}}, \bibinfo {author} {\bibfnamefont {T.}~\bibnamefont {Sakthivel}}, \
  and\ \bibinfo {author} {\bibfnamefont {S.}~\bibnamefont {Seal}},\ }\href@noop
  {} {\bibfield  {journal} {\bibinfo  {journal} {Progress in Materials
  Science}\ }\textbf {\bibinfo {volume} {73}},\ \bibinfo {pages} {44} (\bibinfo
  {year} {2015})}\BibitemShut {NoStop}%
\bibitem [{\citenamefont {Castellanos-Gomez}(2016)}]{Gomez2016}%
  \BibitemOpen
  \bibfield  {author} {\bibinfo {author} {\bibfnamefont {A.}~\bibnamefont
  {Castellanos-Gomez}},\ }\href@noop {} {\bibfield  {journal} {\bibinfo
  {journal} {Nature Photonics}\ }\textbf {\bibinfo {volume} {10}},\ \bibinfo
  {pages} {202} (\bibinfo {year} {2016})}\BibitemShut {NoStop}%
\bibitem [{\citenamefont {Mak}\ and\ \citenamefont {Shan}(2016)}]{Mak2016}%
  \BibitemOpen
  \bibfield  {author} {\bibinfo {author} {\bibfnamefont {K.~F.}\ \bibnamefont
  {Mak}}\ and\ \bibinfo {author} {\bibfnamefont {J.}~\bibnamefont {Shan}},\
  }\href@noop {} {\bibfield  {journal} {\bibinfo  {journal} {Nature Photonics}\
  }\textbf {\bibinfo {volume} {10}},\ \bibinfo {pages} {216} (\bibinfo {year}
  {2016})}\BibitemShut {NoStop}%
\bibitem [{\citenamefont {Mak}\ \emph {et~al.}(2010)\citenamefont {Mak},
  \citenamefont {Lee}, \citenamefont {Hone}, \citenamefont {Shan},\ and\
  \citenamefont {Heinz}}]{MakPRL2010}%
  \BibitemOpen
  \bibfield  {author} {\bibinfo {author} {\bibfnamefont {K.}~\bibnamefont
  {Mak}}, \bibinfo {author} {\bibfnamefont {C.}~\bibnamefont {Lee}}, \bibinfo
  {author} {\bibfnamefont {J.}~\bibnamefont {Hone}}, \bibinfo {author}
  {\bibfnamefont {J.}~\bibnamefont {Shan}}, \ and\ \bibinfo {author}
  {\bibfnamefont {T.}~\bibnamefont {Heinz}},\ }\href@noop {} {\bibfield
  {journal} {\bibinfo  {journal} {Phys. Rev. Lett.}\ }\textbf {\bibinfo
  {volume} {105}},\ \bibinfo {pages} {136805} (\bibinfo {year}
  {2010})}\BibitemShut {NoStop}%
\bibitem [{\citenamefont {Berkelbach}\ \emph {et~al.}(2013)\citenamefont
  {Berkelbach}, \citenamefont {Hybertsen},\ and\ \citenamefont
  {Reichman}}]{Berkelbach2013}%
  \BibitemOpen
  \bibfield  {author} {\bibinfo {author} {\bibfnamefont {T.~C.}\ \bibnamefont
  {Berkelbach}}, \bibinfo {author} {\bibfnamefont {M.~S.}\ \bibnamefont
  {Hybertsen}}, \ and\ \bibinfo {author} {\bibfnamefont {D.~R.}\ \bibnamefont
  {Reichman}},\ }\href {\doibase 10.1103/PhysRevB.88.045318} {\bibfield
  {journal} {\bibinfo  {journal} {Phys. Rev. B}\ }\textbf {\bibinfo {volume}
  {88}},\ \bibinfo {pages} {045318} (\bibinfo {year} {2013})}\BibitemShut
  {NoStop}%
\bibitem [{\citenamefont {Ugeda}\ \emph {et~al.}(2014)\citenamefont {Ugeda},
  \citenamefont {Bradley}, \citenamefont {Shi}, \citenamefont {da~Jornada},
  \citenamefont {Zhang}, \citenamefont {Qiu}, \citenamefont {Ruan},
  \citenamefont {Mo}, \citenamefont {Hussain}, \citenamefont {Shen},
  \citenamefont {Wang}, \citenamefont {Louie},\ and\ \citenamefont
  {Crommie}}]{Ugeda2014}%
  \BibitemOpen
  \bibfield  {author} {\bibinfo {author} {\bibfnamefont {M.~M.}\ \bibnamefont
  {Ugeda}}, \bibinfo {author} {\bibfnamefont {A.~J.}\ \bibnamefont {Bradley}},
  \bibinfo {author} {\bibfnamefont {S.-F.}\ \bibnamefont {Shi}}, \bibinfo
  {author} {\bibfnamefont {F.~H.}\ \bibnamefont {da~Jornada}}, \bibinfo
  {author} {\bibfnamefont {Y.}~\bibnamefont {Zhang}}, \bibinfo {author}
  {\bibfnamefont {D.~Y.}\ \bibnamefont {Qiu}}, \bibinfo {author} {\bibfnamefont
  {W.}~\bibnamefont {Ruan}}, \bibinfo {author} {\bibfnamefont {S.-K.}\
  \bibnamefont {Mo}}, \bibinfo {author} {\bibfnamefont {Z.}~\bibnamefont
  {Hussain}}, \bibinfo {author} {\bibfnamefont {Z.-X.}\ \bibnamefont {Shen}},
  \bibinfo {author} {\bibfnamefont {F.}~\bibnamefont {Wang}}, \bibinfo {author}
  {\bibfnamefont {S.~G.}\ \bibnamefont {Louie}}, \ and\ \bibinfo {author}
  {\bibfnamefont {M.~F.}\ \bibnamefont {Crommie}},\ }\href
  {http://dx.doi.org/10.1038/nmat4061} {\bibfield  {journal} {\bibinfo
  {journal} {Nature Materials}\ }\textbf {\bibinfo {volume} {13}},\ \bibinfo
  {pages} {1091} (\bibinfo {year} {2014})}\BibitemShut {NoStop}%
\bibitem [{\citenamefont {Dufferwiel}\ \emph {et~al.}(2015)\citenamefont
  {Dufferwiel}, \citenamefont {Schwarz}, \citenamefont {Withers}, \citenamefont
  {Trichet}, \citenamefont {Li}, \citenamefont {Sich}, \citenamefont {Del
  Pozo-Zamudio}, \citenamefont {Clark}, \citenamefont {Nalitov}, \citenamefont
  {Solnyshkov}, \citenamefont {Malpuech}, \citenamefont {Novoselov},
  \citenamefont {Smith}, \citenamefont {Skolnick}, \citenamefont
  {Krizhanovskii},\ and\ \citenamefont {Tartakovskii}}]{Dufferwiel2015}%
  \BibitemOpen
  \bibfield  {author} {\bibinfo {author} {\bibfnamefont {S.}~\bibnamefont
  {Dufferwiel}}, \bibinfo {author} {\bibfnamefont {S.}~\bibnamefont {Schwarz}},
  \bibinfo {author} {\bibfnamefont {F.}~\bibnamefont {Withers}}, \bibinfo
  {author} {\bibfnamefont {A.~A.~P.}\ \bibnamefont {Trichet}}, \bibinfo
  {author} {\bibfnamefont {F.}~\bibnamefont {Li}}, \bibinfo {author}
  {\bibfnamefont {M.}~\bibnamefont {Sich}}, \bibinfo {author} {\bibfnamefont
  {O.}~\bibnamefont {Del Pozo-Zamudio}}, \bibinfo {author} {\bibfnamefont
  {C.}~\bibnamefont {Clark}}, \bibinfo {author} {\bibfnamefont
  {A.}~\bibnamefont {Nalitov}}, \bibinfo {author} {\bibfnamefont {D.~D.}\
  \bibnamefont {Solnyshkov}}, \bibinfo {author} {\bibfnamefont
  {G.}~\bibnamefont {Malpuech}}, \bibinfo {author} {\bibfnamefont {K.~S.}\
  \bibnamefont {Novoselov}}, \bibinfo {author} {\bibfnamefont {J.~M.}\
  \bibnamefont {Smith}}, \bibinfo {author} {\bibfnamefont {M.~S.}\ \bibnamefont
  {Skolnick}}, \bibinfo {author} {\bibfnamefont {D.~N.}\ \bibnamefont
  {Krizhanovskii}}, \ and\ \bibinfo {author} {\bibfnamefont {A.~I.}\
  \bibnamefont {Tartakovskii}},\ }\href@noop {} {\bibfield  {journal} {\bibinfo
   {journal} {Nat Commun}\ }\textbf {\bibinfo {volume} {6}},\ \bibinfo {pages}
  {8579} (\bibinfo {year} {2015})}\BibitemShut {NoStop}%
\bibitem [{\citenamefont {Sidler}\ \emph {et~al.}(2016)\citenamefont {Sidler},
  \citenamefont {Back}, \citenamefont {Cotlet}, \citenamefont {Srivastava},
  \citenamefont {Fink}, \citenamefont {Kroner}, \citenamefont {Demler},\ and\
  \citenamefont {Imamoglu}}]{Sidler2016}%
  \BibitemOpen
  \bibfield  {author} {\bibinfo {author} {\bibfnamefont {M.}~\bibnamefont
  {Sidler}}, \bibinfo {author} {\bibfnamefont {P.}~\bibnamefont {Back}},
  \bibinfo {author} {\bibfnamefont {O.}~\bibnamefont {Cotlet}}, \bibinfo
  {author} {\bibfnamefont {A.}~\bibnamefont {Srivastava}}, \bibinfo {author}
  {\bibfnamefont {T.}~\bibnamefont {Fink}}, \bibinfo {author} {\bibfnamefont
  {M.}~\bibnamefont {Kroner}}, \bibinfo {author} {\bibfnamefont
  {E.}~\bibnamefont {Demler}}, \ and\ \bibinfo {author} {\bibfnamefont
  {A.}~\bibnamefont {Imamoglu}},\ }\href {\doibase 10.1038/nphys3949}
  {\bibfield  {journal} {\bibinfo  {journal} {Nature Physics}\ } (\bibinfo
  {year} {2016}),\ 10.1038/nphys3949}\BibitemShut {NoStop}%
\bibitem [{\citenamefont {Liu}\ \emph {et~al.}(2015)\citenamefont {Liu},
  \citenamefont {Galfsky}, \citenamefont {Sun}, \citenamefont {Xia},
  \citenamefont {Lin}, \citenamefont {Lee}, \citenamefont {K{\'e}na-Cohen},\
  and\ \citenamefont {Menon}}]{Liu2015}%
  \BibitemOpen
  \bibfield  {author} {\bibinfo {author} {\bibfnamefont {X.}~\bibnamefont
  {Liu}}, \bibinfo {author} {\bibfnamefont {T.}~\bibnamefont {Galfsky}},
  \bibinfo {author} {\bibfnamefont {Z.}~\bibnamefont {Sun}}, \bibinfo {author}
  {\bibfnamefont {F.}~\bibnamefont {Xia}}, \bibinfo {author} {\bibfnamefont
  {E.-c.}\ \bibnamefont {Lin}}, \bibinfo {author} {\bibfnamefont {Y.-H.}\
  \bibnamefont {Lee}}, \bibinfo {author} {\bibfnamefont {S.}~\bibnamefont
  {K{\'e}na-Cohen}}, \ and\ \bibinfo {author} {\bibfnamefont {V.~M.}\
  \bibnamefont {Menon}},\ }\href {http://dx.doi.org/10.1038/nphoton.2014.304}
  {\bibfield  {journal} {\bibinfo  {journal} {Nat Photon}\ }\textbf {\bibinfo
  {volume} {9}},\ \bibinfo {pages} {30} (\bibinfo {year} {2015})}\BibitemShut
  {NoStop}%
\bibitem [{\citenamefont {Flatten}\ \emph {et~al.}(2016)\citenamefont
  {Flatten}, \citenamefont {He}, \citenamefont {Coles}, \citenamefont
  {Trichet}, \citenamefont {Powell}, \citenamefont {Taylor}, \citenamefont
  {Warner},\ and\ \citenamefont {Smith}}]{Flatten2016}%
  \BibitemOpen
  \bibfield  {author} {\bibinfo {author} {\bibfnamefont {L.~C.}\ \bibnamefont
  {Flatten}}, \bibinfo {author} {\bibfnamefont {Z.}~\bibnamefont {He}},
  \bibinfo {author} {\bibfnamefont {D.~M.}\ \bibnamefont {Coles}}, \bibinfo
  {author} {\bibfnamefont {A.~A.~P.}\ \bibnamefont {Trichet}}, \bibinfo
  {author} {\bibfnamefont {A.~W.}\ \bibnamefont {Powell}}, \bibinfo {author}
  {\bibfnamefont {R.~A.}\ \bibnamefont {Taylor}}, \bibinfo {author}
  {\bibfnamefont {J.~H.}\ \bibnamefont {Warner}}, \ and\ \bibinfo {author}
  {\bibfnamefont {J.~M.}\ \bibnamefont {Smith}},\ }\href@noop {} {\bibfield
  {journal} {\bibinfo  {journal} {Scientific Reports}\ }\textbf {\bibinfo
  {volume} {6}},\ \bibinfo {pages} {33134} (\bibinfo {year}
  {2016})}\BibitemShut {NoStop}%
\bibitem [{\citenamefont {Lundt}\ \emph {et~al.}(2016)\citenamefont {Lundt},
  \citenamefont {Maynski}, \citenamefont {Cherotchenko}, \citenamefont {Pant},
  \citenamefont {Fan}, \citenamefont {Sek}, \citenamefont {Tongay},
  \citenamefont {Kavokin}, \citenamefont {Hofling},\ and\ \citenamefont
  {Schneider}}]{Lundt2016}%
  \BibitemOpen
  \bibfield  {author} {\bibinfo {author} {\bibfnamefont {N.}~\bibnamefont
  {Lundt}}, \bibinfo {author} {\bibfnamefont {A.}~\bibnamefont {Maynski}},
  \bibinfo {author} {\bibfnamefont {E.}~\bibnamefont {Cherotchenko}}, \bibinfo
  {author} {\bibfnamefont {A.}~\bibnamefont {Pant}}, \bibinfo {author}
  {\bibfnamefont {X.}~\bibnamefont {Fan}}, \bibinfo {author} {\bibfnamefont
  {G.}~\bibnamefont {Sek}}, \bibinfo {author} {\bibfnamefont {S.}~\bibnamefont
  {Tongay}}, \bibinfo {author} {\bibfnamefont {A.~V.}\ \bibnamefont {Kavokin}},
  \bibinfo {author} {\bibfnamefont {S.}~\bibnamefont {Hofling}}, \ and\
  \bibinfo {author} {\bibfnamefont {C.}~\bibnamefont {Schneider}},\ }\href@noop
  {} {\bibfield  {journal} {\bibinfo  {journal}
  {http://arxiv.org/abs/1603.05562}\ } (\bibinfo {year} {2016})}\BibitemShut
  {NoStop}%
\bibitem [{\citenamefont {Mak}\ \emph {et~al.}(2014)\citenamefont {Mak},
  \citenamefont {McGill}, \citenamefont {Park},\ and\ \citenamefont
  {McEuen}}]{Mak2014}%
  \BibitemOpen
  \bibfield  {author} {\bibinfo {author} {\bibfnamefont {K.~F.}\ \bibnamefont
  {Mak}}, \bibinfo {author} {\bibfnamefont {K.~L.}\ \bibnamefont {McGill}},
  \bibinfo {author} {\bibfnamefont {J.}~\bibnamefont {Park}}, \ and\ \bibinfo
  {author} {\bibfnamefont {P.~L.}\ \bibnamefont {McEuen}},\ }\href@noop {}
  {\bibfield  {journal} {\bibinfo  {journal} {Science}\ }\textbf {\bibinfo
  {volume} {344}},\ \bibinfo {pages} {1489} (\bibinfo {year}
  {2014})}\BibitemShut {NoStop}%
\bibitem [{\citenamefont {Xiao}\ \emph {et~al.}(2007)\citenamefont {Xiao},
  \citenamefont {Yao},\ and\ \citenamefont {Niu}}]{Xiao2007}%
  \BibitemOpen
  \bibfield  {author} {\bibinfo {author} {\bibfnamefont {D.}~\bibnamefont
  {Xiao}}, \bibinfo {author} {\bibfnamefont {W.}~\bibnamefont {Yao}}, \ and\
  \bibinfo {author} {\bibfnamefont {Q.}~\bibnamefont {Niu}},\ }\href@noop {}
  {\bibfield  {journal} {\bibinfo  {journal} {Phys. Rev. Letters}\ }\textbf
  {\bibinfo {volume} {99}},\ \bibinfo {pages} {236809} (\bibinfo {year}
  {2007})}\BibitemShut {NoStop}%
\bibitem [{\citenamefont {Chang}\ and\ \citenamefont {Niu}(2008)}]{Chang2008}%
  \BibitemOpen
  \bibfield  {author} {\bibinfo {author} {\bibfnamefont {M.-C.}\ \bibnamefont
  {Chang}}\ and\ \bibinfo {author} {\bibfnamefont {Q.}~\bibnamefont {Niu}},\
  }\href@noop {} {\bibfield  {journal} {\bibinfo  {journal} {J. Phys.: Condens.
  Matter}\ }\textbf {\bibinfo {volume} {20}},\ \bibinfo {pages} {193202}
  (\bibinfo {year} {2008})}\BibitemShut {NoStop}%
\bibitem [{\citenamefont {Mak}\ \emph {et~al.}(2012)\citenamefont {Mak},
  \citenamefont {He}, \citenamefont {Shan},\ and\ \citenamefont
  {Heinz}}]{Mak2012}%
  \BibitemOpen
  \bibfield  {author} {\bibinfo {author} {\bibfnamefont {K.~F.}\ \bibnamefont
  {Mak}}, \bibinfo {author} {\bibfnamefont {K.}~\bibnamefont {He}}, \bibinfo
  {author} {\bibfnamefont {J.}~\bibnamefont {Shan}}, \ and\ \bibinfo {author}
  {\bibfnamefont {T.~F.}\ \bibnamefont {Heinz}},\ }\href {\doibase
  10.1038/nnano.2012.96} {\bibfield  {journal} {\bibinfo  {journal} {Nature
  nanotechnology}\ }\textbf {\bibinfo {volume} {7}},\ \bibinfo {pages} {494}
  (\bibinfo {year} {2012})}\BibitemShut {NoStop}%
\bibitem [{\citenamefont {Rycerz}\ \emph {et~al.}(2007)\citenamefont {Rycerz},
  \citenamefont {Tworzydlo},\ and\ \citenamefont {Beenakker}}]{Beenakker2007}%
  \BibitemOpen
  \bibfield  {author} {\bibinfo {author} {\bibfnamefont {A.}~\bibnamefont
  {Rycerz}}, \bibinfo {author} {\bibfnamefont {J.}~\bibnamefont {Tworzydlo}}, \
  and\ \bibinfo {author} {\bibfnamefont {C.~W.~J.}\ \bibnamefont {Beenakker}},\
  }\href@noop {} {\bibfield  {journal} {\bibinfo  {journal} {Nature Physics}\
  }\textbf {\bibinfo {volume} {3}},\ \bibinfo {pages} {172} (\bibinfo {year}
  {2007})}\BibitemShut {NoStop}%
\bibitem [{\citenamefont {Jones}\ \emph {et~al.}(2013)\citenamefont {Jones},
  \citenamefont {Yu}, \citenamefont {Ghimire}, \citenamefont {Wu},
  \citenamefont {Aivazian}, \citenamefont {Ross}, \citenamefont {Zhao},
  \citenamefont {Yan}, \citenamefont {Mandrus}, \citenamefont {Xiao},
  \citenamefont {Yao},\ and\ \citenamefont {Xu}}]{Jones2013}%
  \BibitemOpen
  \bibfield  {author} {\bibinfo {author} {\bibfnamefont {A.~M.}\ \bibnamefont
  {Jones}}, \bibinfo {author} {\bibfnamefont {H.}~\bibnamefont {Yu}}, \bibinfo
  {author} {\bibfnamefont {N.~J.}\ \bibnamefont {Ghimire}}, \bibinfo {author}
  {\bibfnamefont {S.}~\bibnamefont {Wu}}, \bibinfo {author} {\bibfnamefont
  {G.}~\bibnamefont {Aivazian}}, \bibinfo {author} {\bibfnamefont {J.~S.}\
  \bibnamefont {Ross}}, \bibinfo {author} {\bibfnamefont {B.}~\bibnamefont
  {Zhao}}, \bibinfo {author} {\bibfnamefont {J.}~\bibnamefont {Yan}}, \bibinfo
  {author} {\bibfnamefont {D.~G.}\ \bibnamefont {Mandrus}}, \bibinfo {author}
  {\bibfnamefont {D.}~\bibnamefont {Xiao}}, \bibinfo {author} {\bibfnamefont
  {W.}~\bibnamefont {Yao}}, \ and\ \bibinfo {author} {\bibfnamefont
  {X.}~\bibnamefont {Xu}},\ }\href {http://dx.doi.org/10.1038/nnano.2013.151}
  {\bibfield  {journal} {\bibinfo  {journal} {Nat Nano}\ }\textbf {\bibinfo
  {volume} {8}},\ \bibinfo {pages} {634} (\bibinfo {year} {2013})}\BibitemShut
  {NoStop}%
\bibitem [{\citenamefont {Suzuki}\ \emph {et~al.}(2014)\citenamefont {Suzuki},
  \citenamefont {Sakano}, \citenamefont {Zhang}, \citenamefont {Akashi},
  \citenamefont {Morikawa}, \citenamefont {Harasawa}, \citenamefont {Yaji},
  \citenamefont {Kuroda}, \citenamefont {Miyamoto}, \citenamefont {Okuda},
  \citenamefont {Ishizaka}, \citenamefont {Arita},\ and\ \citenamefont
  {Iwasa}}]{SuzukiR2014}%
  \BibitemOpen
  \bibfield  {author} {\bibinfo {author} {\bibfnamefont {R.}~\bibnamefont
  {Suzuki}}, \bibinfo {author} {\bibfnamefont {M.}~\bibnamefont {Sakano}},
  \bibinfo {author} {\bibfnamefont {Y.~J.}\ \bibnamefont {Zhang}}, \bibinfo
  {author} {\bibfnamefont {R.}~\bibnamefont {Akashi}}, \bibinfo {author}
  {\bibfnamefont {D.}~\bibnamefont {Morikawa}}, \bibinfo {author}
  {\bibfnamefont {A.}~\bibnamefont {Harasawa}}, \bibinfo {author}
  {\bibfnamefont {K.}~\bibnamefont {Yaji}}, \bibinfo {author} {\bibfnamefont
  {K.}~\bibnamefont {Kuroda}}, \bibinfo {author} {\bibfnamefont
  {K.}~\bibnamefont {Miyamoto}}, \bibinfo {author} {\bibfnamefont
  {T.}~\bibnamefont {Okuda}}, \bibinfo {author} {\bibfnamefont
  {K.}~\bibnamefont {Ishizaka}}, \bibinfo {author} {\bibfnamefont
  {R.}~\bibnamefont {Arita}}, \ and\ \bibinfo {author} {\bibfnamefont
  {Y.}~\bibnamefont {Iwasa}},\ }\href
  {http://dx.doi.org/10.1038/nnano.2014.148} {\bibfield  {journal} {\bibinfo
  {journal} {Nat Nano}\ }\textbf {\bibinfo {volume} {9}},\ \bibinfo {pages}
  {611} (\bibinfo {year} {2014})}\BibitemShut {NoStop}%
\bibitem [{\citenamefont {Zhu}\ \emph {et~al.}(2014)\citenamefont {Zhu},
  \citenamefont {Zeng}, \citenamefont {Dai}, \citenamefont {Gong},\ and\
  \citenamefont {Cui}}]{Zhu2014}%
  \BibitemOpen
  \bibfield  {author} {\bibinfo {author} {\bibfnamefont {B.}~\bibnamefont
  {Zhu}}, \bibinfo {author} {\bibfnamefont {H.}~\bibnamefont {Zeng}}, \bibinfo
  {author} {\bibfnamefont {J.}~\bibnamefont {Dai}}, \bibinfo {author}
  {\bibfnamefont {Z.}~\bibnamefont {Gong}}, \ and\ \bibinfo {author}
  {\bibfnamefont {X.}~\bibnamefont {Cui}},\ }\href@noop {} {\bibfield
  {journal} {\bibinfo  {journal} {Proceedings of the National Academy of
  Sciences}\ }\textbf {\bibinfo {volume} {111}},\ \bibinfo {pages} {11606}
  (\bibinfo {year} {2014})}\BibitemShut {NoStop}%
\bibitem [{\citenamefont {Sallen}\ \emph {et~al.}(2012)\citenamefont {Sallen},
  \citenamefont {Bouet}, \citenamefont {Marie}, \citenamefont {Wang},
  \citenamefont {Zhu}, \citenamefont {Han}, \citenamefont {Lu}, \citenamefont
  {Tan}, \citenamefont {Amand}, \citenamefont {Liu},\ and\ \citenamefont
  {Urbaszek}}]{Sallen2012}%
  \BibitemOpen
  \bibfield  {author} {\bibinfo {author} {\bibfnamefont {G.}~\bibnamefont
  {Sallen}}, \bibinfo {author} {\bibfnamefont {L.}~\bibnamefont {Bouet}},
  \bibinfo {author} {\bibfnamefont {X.}~\bibnamefont {Marie}}, \bibinfo
  {author} {\bibfnamefont {G.}~\bibnamefont {Wang}}, \bibinfo {author}
  {\bibfnamefont {C.~R.}\ \bibnamefont {Zhu}}, \bibinfo {author} {\bibfnamefont
  {W.~P.}\ \bibnamefont {Han}}, \bibinfo {author} {\bibfnamefont
  {Y.}~\bibnamefont {Lu}}, \bibinfo {author} {\bibfnamefont {P.~H.}\
  \bibnamefont {Tan}}, \bibinfo {author} {\bibfnamefont {T.}~\bibnamefont
  {Amand}}, \bibinfo {author} {\bibfnamefont {B.~L.}\ \bibnamefont {Liu}}, \
  and\ \bibinfo {author} {\bibfnamefont {B.}~\bibnamefont {Urbaszek}},\ }\href
  {\doibase 10.1103/PhysRevB.86.081301} {\bibfield  {journal} {\bibinfo
  {journal} {Phys. Rev. B}\ }\textbf {\bibinfo {volume} {86}},\ \bibinfo
  {pages} {081301} (\bibinfo {year} {2012})}\BibitemShut {NoStop}%
\bibitem [{\citenamefont {Kioseoglou}\ \emph {et~al.}(2012)\citenamefont
  {Kioseoglou}, \citenamefont {Hanbicki}, \citenamefont {Currie}, \citenamefont
  {Friedman}, \citenamefont {Gunlycke},\ and\ \citenamefont
  {Jonker}}]{Kioseoglou2012}%
  \BibitemOpen
  \bibfield  {author} {\bibinfo {author} {\bibfnamefont {G.}~\bibnamefont
  {Kioseoglou}}, \bibinfo {author} {\bibfnamefont {A.~T.}\ \bibnamefont
  {Hanbicki}}, \bibinfo {author} {\bibfnamefont {M.}~\bibnamefont {Currie}},
  \bibinfo {author} {\bibfnamefont {A.~L.}\ \bibnamefont {Friedman}}, \bibinfo
  {author} {\bibfnamefont {D.}~\bibnamefont {Gunlycke}}, \ and\ \bibinfo
  {author} {\bibfnamefont {B.~T.}\ \bibnamefont {Jonker}},\ }\href@noop {}
  {\bibfield  {journal} {\bibinfo  {journal} {Applied Physics Letters}\
  }\textbf {\bibinfo {volume} {101}} (\bibinfo {year} {2012})}\BibitemShut
  {NoStop}%
\bibitem [{\citenamefont {Lagarde}\ \emph {et~al.}(2014)\citenamefont
  {Lagarde}, \citenamefont {Bouet}, \citenamefont {Marie}, \citenamefont {Zhu},
  \citenamefont {Liu}, \citenamefont {Amand}, \citenamefont {Tan},\ and\
  \citenamefont {Urbaszek}}]{Lagarde2014}%
  \BibitemOpen
  \bibfield  {author} {\bibinfo {author} {\bibfnamefont {D.}~\bibnamefont
  {Lagarde}}, \bibinfo {author} {\bibfnamefont {L.}~\bibnamefont {Bouet}},
  \bibinfo {author} {\bibfnamefont {X.}~\bibnamefont {Marie}}, \bibinfo
  {author} {\bibfnamefont {C.~R.}\ \bibnamefont {Zhu}}, \bibinfo {author}
  {\bibfnamefont {B.~L.}\ \bibnamefont {Liu}}, \bibinfo {author} {\bibfnamefont
  {T.}~\bibnamefont {Amand}}, \bibinfo {author} {\bibfnamefont {P.~H.}\
  \bibnamefont {Tan}}, \ and\ \bibinfo {author} {\bibfnamefont
  {B.}~\bibnamefont {Urbaszek}},\ }\href {\doibase
  10.1103/PhysRevLett.112.047401} {\bibfield  {journal} {\bibinfo  {journal}
  {Phys. Rev. Lett.}\ }\textbf {\bibinfo {volume} {112}},\ \bibinfo {pages}
  {047401} (\bibinfo {year} {2014})}\BibitemShut {NoStop}%
\bibitem [{\citenamefont {Yu}\ \emph {et~al.}(2014)\citenamefont {Yu},
  \citenamefont {Liu}, \citenamefont {Gong}, \citenamefont {Xu},\ and\
  \citenamefont {Yao}}]{Yu2014}%
  \BibitemOpen
  \bibfield  {author} {\bibinfo {author} {\bibfnamefont {H.}~\bibnamefont
  {Yu}}, \bibinfo {author} {\bibfnamefont {G.-B.}\ \bibnamefont {Liu}},
  \bibinfo {author} {\bibfnamefont {P.}~\bibnamefont {Gong}}, \bibinfo {author}
  {\bibfnamefont {X.}~\bibnamefont {Xu}}, \ and\ \bibinfo {author}
  {\bibfnamefont {W.}~\bibnamefont {Yao}},\ }\href@noop {} {\bibfield
  {journal} {\bibinfo  {journal} {Nat Commun}\ }\textbf {\bibinfo {volume} {5}}
  (\bibinfo {year} {2014})}\BibitemShut {NoStop}%
\bibitem [{\citenamefont {Glazov}\ \emph {et~al.}(2014)\citenamefont {Glazov},
  \citenamefont {Amand}, \citenamefont {Marie}, \citenamefont {Lagarde},
  \citenamefont {Bouet},\ and\ \citenamefont {Urbaszek}}]{Glazov2014}%
  \BibitemOpen
  \bibfield  {author} {\bibinfo {author} {\bibfnamefont {M.~M.}\ \bibnamefont
  {Glazov}}, \bibinfo {author} {\bibfnamefont {T.}~\bibnamefont {Amand}},
  \bibinfo {author} {\bibfnamefont {X.}~\bibnamefont {Marie}}, \bibinfo
  {author} {\bibfnamefont {D.}~\bibnamefont {Lagarde}}, \bibinfo {author}
  {\bibfnamefont {L.}~\bibnamefont {Bouet}}, \ and\ \bibinfo {author}
  {\bibfnamefont {B.}~\bibnamefont {Urbaszek}},\ }\href {\doibase
  10.1103/PhysRevB.89.201302} {\bibfield  {journal} {\bibinfo  {journal} {Phys.
  Rev. B}\ }\textbf {\bibinfo {volume} {89}},\ \bibinfo {pages} {201302}
  (\bibinfo {year} {2014})}\BibitemShut {NoStop}%
\bibitem [{\citenamefont {Maialle}\ \emph {et~al.}(1993)\citenamefont
  {Maialle}, \citenamefont {de~Andrada~e Silva},\ and\ \citenamefont
  {Sham}}]{Maialle1993}%
  \BibitemOpen
  \bibfield  {author} {\bibinfo {author} {\bibfnamefont {M.~Z.}\ \bibnamefont
  {Maialle}}, \bibinfo {author} {\bibfnamefont {E.~A.}\ \bibnamefont
  {de~Andrada~e Silva}}, \ and\ \bibinfo {author} {\bibfnamefont {L.~J.}\
  \bibnamefont {Sham}},\ }\href {\doibase 10.1103/PhysRevB.47.15776} {\bibfield
   {journal} {\bibinfo  {journal} {Phys. Rev. B}\ }\textbf {\bibinfo {volume}
  {47}},\ \bibinfo {pages} {15776} (\bibinfo {year} {1993})}\BibitemShut
  {NoStop}%
\bibitem [{\citenamefont {Dyakonov}\ and\ \citenamefont
  {Perel}(1972)}]{Dyakonov2}%
  \BibitemOpen
  \bibfield  {author} {\bibinfo {author} {\bibfnamefont {M.}~\bibnamefont
  {Dyakonov}}\ and\ \bibinfo {author} {\bibfnamefont {V.}~\bibnamefont
  {Perel}},\ }\href@noop {} {\bibfield  {journal} {\bibinfo  {journal} {Sov.
  Phys. Solid State}\ }\textbf {\bibinfo {volume} {13}},\ \bibinfo {pages}
  {3023} (\bibinfo {year} {1972})}\BibitemShut {NoStop}%
\bibitem [{\citenamefont {Kavokin}\ \emph {et~al.}(2005)\citenamefont
  {Kavokin}, \citenamefont {Malpuech},\ and\ \citenamefont
  {Glazov}}]{Kavokin2005}%
  \BibitemOpen
  \bibfield  {author} {\bibinfo {author} {\bibfnamefont {A.}~\bibnamefont
  {Kavokin}}, \bibinfo {author} {\bibfnamefont {G.}~\bibnamefont {Malpuech}}, \
  and\ \bibinfo {author} {\bibfnamefont {M.}~\bibnamefont {Glazov}},\
  }\href@noop {} {\bibfield  {journal} {\bibinfo  {journal} {Phys. Rev. Lett.}\
  }\textbf {\bibinfo {volume} {95}},\ \bibinfo {pages} {136601} (\bibinfo
  {year} {2005})}\BibitemShut {NoStop}%
\bibitem [{\citenamefont {Kavokin}\ \emph {et~al.}(2011)\citenamefont
  {Kavokin}, \citenamefont {Baumberg}, \citenamefont {Malpuech},\ and\
  \citenamefont {Laussy}}]{Microcavities}%
  \BibitemOpen
  \bibfield  {author} {\bibinfo {author} {\bibfnamefont {A.}~\bibnamefont
  {Kavokin}}, \bibinfo {author} {\bibfnamefont {J.~J.}\ \bibnamefont
  {Baumberg}}, \bibinfo {author} {\bibfnamefont {G.}~\bibnamefont {Malpuech}},
  \ and\ \bibinfo {author} {\bibfnamefont {F.~P.}\ \bibnamefont {Laussy}},\
  }\href@noop {} {\emph {\bibinfo {title} {Microcavities}}}\ (\bibinfo
  {publisher} {Oxford University Press},\ \bibinfo {year} {2011})\BibitemShut
  {NoStop}%
\bibitem [{\citenamefont {Panzarini}\ \emph {et~al.}(1999)\citenamefont
  {Panzarini}, \citenamefont {Andreani}, \citenamefont {Armitage},
  \citenamefont {Baxter}, \citenamefont {Skolnick}, \citenamefont {Astratov},
  \citenamefont {Roberts}, \citenamefont {Kavokin}, \citenamefont
  {Vladimirova},\ and\ \citenamefont {Kaliteevski}}]{Panzarini99}%
  \BibitemOpen
  \bibfield  {author} {\bibinfo {author} {\bibfnamefont {G.}~\bibnamefont
  {Panzarini}}, \bibinfo {author} {\bibfnamefont {L.}~\bibnamefont {Andreani}},
  \bibinfo {author} {\bibfnamefont {A.}~\bibnamefont {Armitage}}, \bibinfo
  {author} {\bibfnamefont {D.}~\bibnamefont {Baxter}}, \bibinfo {author}
  {\bibfnamefont {M.}~\bibnamefont {Skolnick}}, \bibinfo {author}
  {\bibfnamefont {V.}~\bibnamefont {Astratov}}, \bibinfo {author}
  {\bibfnamefont {J.}~\bibnamefont {Roberts}}, \bibinfo {author} {\bibfnamefont
  {A.}~\bibnamefont {Kavokin}}, \bibinfo {author} {\bibfnamefont
  {M.}~\bibnamefont {Vladimirova}}, \ and\ \bibinfo {author} {\bibfnamefont
  {M.}~\bibnamefont {Kaliteevski}},\ }\href@noop {} {\bibfield  {journal}
  {\bibinfo  {journal} {Phys. Rev. B}\ }\textbf {\bibinfo {volume} {59}},\
  \bibinfo {pages} {5082} (\bibinfo {year} {1999})}\BibitemShut {NoStop}%
\bibitem [{\citenamefont {Shelykh}\ \emph {et~al.}(2010)\citenamefont
  {Shelykh}, \citenamefont {Kavokin}, \citenamefont {Rubo}, \citenamefont
  {Liew},\ and\ \citenamefont {Malpuech}}]{Shelykh2010}%
  \BibitemOpen
  \bibfield  {author} {\bibinfo {author} {\bibfnamefont {I.~A.}\ \bibnamefont
  {Shelykh}}, \bibinfo {author} {\bibfnamefont {A.~V.}\ \bibnamefont
  {Kavokin}}, \bibinfo {author} {\bibfnamefont {Y.~G.}\ \bibnamefont {Rubo}},
  \bibinfo {author} {\bibfnamefont {T.~C.~H.}\ \bibnamefont {Liew}}, \ and\
  \bibinfo {author} {\bibfnamefont {G.}~\bibnamefont {Malpuech}},\ }\href
  {http://stacks.iop.org/0268-1242/25/i=1/a=013001} {\bibfield  {journal}
  {\bibinfo  {journal} {Semiconductor Science and Technology}\ }\textbf
  {\bibinfo {volume} {25}},\ \bibinfo {pages} {013001} (\bibinfo {year}
  {2010})}\BibitemShut {NoStop}%
\bibitem [{\citenamefont {Leyder}\ \emph {et~al.}(2007)\citenamefont {Leyder},
  \citenamefont {Romanelli}, \citenamefont {Karr}, \citenamefont {Giacobino},
  \citenamefont {Liew}, \citenamefont {Glazov}, \citenamefont {Kavokin},
  \citenamefont {Malpuech},\ and\ \citenamefont {Bramati}}]{Leyder2007}%
  \BibitemOpen
  \bibfield  {author} {\bibinfo {author} {\bibfnamefont {C.}~\bibnamefont
  {Leyder}}, \bibinfo {author} {\bibfnamefont {M.}~\bibnamefont {Romanelli}},
  \bibinfo {author} {\bibfnamefont {J.~P.}\ \bibnamefont {Karr}}, \bibinfo
  {author} {\bibfnamefont {E.}~\bibnamefont {Giacobino}}, \bibinfo {author}
  {\bibfnamefont {T.~C.~H.}\ \bibnamefont {Liew}}, \bibinfo {author}
  {\bibfnamefont {M.~M.}\ \bibnamefont {Glazov}}, \bibinfo {author}
  {\bibfnamefont {A.~V.}\ \bibnamefont {Kavokin}}, \bibinfo {author}
  {\bibfnamefont {G.}~\bibnamefont {Malpuech}}, \ and\ \bibinfo {author}
  {\bibfnamefont {A.}~\bibnamefont {Bramati}},\ }\href
  {http://dx.doi.org/10.1038/nphys676} {\bibfield  {journal} {\bibinfo
  {journal} {Nat Phys}\ }\textbf {\bibinfo {volume} {3}},\ \bibinfo {pages}
  {628} (\bibinfo {year} {2007})}\BibitemShut {NoStop}%
\bibitem [{\citenamefont {Maragkou}\ \emph {et~al.}(2011)\citenamefont
  {Maragkou}, \citenamefont {Richards}, \citenamefont {Ostatnick{\`y}},
  \citenamefont {Grundy}, \citenamefont {Zajac}, \citenamefont {Hugues},
  \citenamefont {Langbein},\ and\ \citenamefont
  {Lagoudakis}}]{maragkou2011optical}%
  \BibitemOpen
  \bibfield  {author} {\bibinfo {author} {\bibfnamefont {M.}~\bibnamefont
  {Maragkou}}, \bibinfo {author} {\bibfnamefont {C.~E.}\ \bibnamefont
  {Richards}}, \bibinfo {author} {\bibfnamefont {T.}~\bibnamefont
  {Ostatnick{\`y}}}, \bibinfo {author} {\bibfnamefont {A.~J.}\ \bibnamefont
  {Grundy}}, \bibinfo {author} {\bibfnamefont {J.}~\bibnamefont {Zajac}},
  \bibinfo {author} {\bibfnamefont {M.}~\bibnamefont {Hugues}}, \bibinfo
  {author} {\bibfnamefont {W.}~\bibnamefont {Langbein}}, \ and\ \bibinfo
  {author} {\bibfnamefont {P.~G.}\ \bibnamefont {Lagoudakis}},\ }\href@noop {}
  {\bibfield  {journal} {\bibinfo  {journal} {Optics letters}\ }\textbf
  {\bibinfo {volume} {36}},\ \bibinfo {pages} {1095} (\bibinfo {year}
  {2011})}\BibitemShut {NoStop}%
\bibitem [{\citenamefont {Kammann}\ \emph {et~al.}(2012)\citenamefont
  {Kammann}, \citenamefont {Liew}, \citenamefont {Ohadi}, \citenamefont
  {Cilibrizzi}, \citenamefont {Tsotsis}, \citenamefont {Hatzopoulos},
  \citenamefont {Savvidis}, \citenamefont {Kavokin},\ and\ \citenamefont
  {Lagoudakis}}]{Lagoudakis2012}%
  \BibitemOpen
  \bibfield  {author} {\bibinfo {author} {\bibfnamefont {E.}~\bibnamefont
  {Kammann}}, \bibinfo {author} {\bibfnamefont {T.~C.~H.}\ \bibnamefont
  {Liew}}, \bibinfo {author} {\bibfnamefont {H.}~\bibnamefont {Ohadi}},
  \bibinfo {author} {\bibfnamefont {P.}~\bibnamefont {Cilibrizzi}}, \bibinfo
  {author} {\bibfnamefont {P.}~\bibnamefont {Tsotsis}}, \bibinfo {author}
  {\bibfnamefont {Z.}~\bibnamefont {Hatzopoulos}}, \bibinfo {author}
  {\bibfnamefont {P.~G.}\ \bibnamefont {Savvidis}}, \bibinfo {author}
  {\bibfnamefont {A.~V.}\ \bibnamefont {Kavokin}}, \ and\ \bibinfo {author}
  {\bibfnamefont {P.~G.}\ \bibnamefont {Lagoudakis}},\ }\href {\doibase
  10.1103/PhysRevLett.109.036404} {\bibfield  {journal} {\bibinfo  {journal}
  {Phys. Rev. Lett.}\ }\textbf {\bibinfo {volume} {109}},\ \bibinfo {pages}
  {036404} (\bibinfo {year} {2012})}\BibitemShut {NoStop}%
\bibitem [{\citenamefont {Sala}\ \emph {et~al.}(2015)\citenamefont {Sala},
  \citenamefont {Solnyshkov}, \citenamefont {Carusotto}, \citenamefont
  {Jacqmin}, \citenamefont {Lema\^{\i}tre}, \citenamefont
  {Ter\ifmmode~\mbox{\c{c}}\else \c{c}\fi{}as}, \citenamefont {Nalitov},
  \citenamefont {Abbarchi}, \citenamefont {Galopin}, \citenamefont {Sagnes},
  \citenamefont {Bloch}, \citenamefont {Malpuech},\ and\ \citenamefont
  {Amo}}]{Sala2014}%
  \BibitemOpen
  \bibfield  {author} {\bibinfo {author} {\bibfnamefont {V.~G.}\ \bibnamefont
  {Sala}}, \bibinfo {author} {\bibfnamefont {D.~D.}\ \bibnamefont
  {Solnyshkov}}, \bibinfo {author} {\bibfnamefont {I.}~\bibnamefont
  {Carusotto}}, \bibinfo {author} {\bibfnamefont {T.}~\bibnamefont {Jacqmin}},
  \bibinfo {author} {\bibfnamefont {A.}~\bibnamefont {Lema\^{\i}tre}}, \bibinfo
  {author} {\bibfnamefont {H.}~\bibnamefont {Ter\ifmmode~\mbox{\c{c}}\else
  \c{c}\fi{}as}}, \bibinfo {author} {\bibfnamefont {A.}~\bibnamefont
  {Nalitov}}, \bibinfo {author} {\bibfnamefont {M.}~\bibnamefont {Abbarchi}},
  \bibinfo {author} {\bibfnamefont {E.}~\bibnamefont {Galopin}}, \bibinfo
  {author} {\bibfnamefont {I.}~\bibnamefont {Sagnes}}, \bibinfo {author}
  {\bibfnamefont {J.}~\bibnamefont {Bloch}}, \bibinfo {author} {\bibfnamefont
  {G.}~\bibnamefont {Malpuech}}, \ and\ \bibinfo {author} {\bibfnamefont
  {A.}~\bibnamefont {Amo}},\ }\href {\doibase 10.1103/PhysRevX.5.011034}
  {\bibfield  {journal} {\bibinfo  {journal} {Phys. Rev. X}\ }\textbf {\bibinfo
  {volume} {5}},\ \bibinfo {pages} {011034} (\bibinfo {year}
  {2015})}\BibitemShut {NoStop}%
\bibitem [{\citenamefont {Nalitov}\ \emph {et~al.}(2015)\citenamefont
  {Nalitov}, \citenamefont {Solnyshkov},\ and\ \citenamefont
  {Malpuech}}]{Nalitov2014b}%
  \BibitemOpen
  \bibfield  {author} {\bibinfo {author} {\bibfnamefont {A.~V.}\ \bibnamefont
  {Nalitov}}, \bibinfo {author} {\bibfnamefont {D.~D.}\ \bibnamefont
  {Solnyshkov}}, \ and\ \bibinfo {author} {\bibfnamefont {G.}~\bibnamefont
  {Malpuech}},\ }\href {\doibase 10.1103/PhysRevLett.114.116401} {\bibfield
  {journal} {\bibinfo  {journal} {Phys. Rev. Lett.}\ }\textbf {\bibinfo
  {volume} {114}},\ \bibinfo {pages} {116401} (\bibinfo {year}
  {2015})}\BibitemShut {NoStop}%
\bibitem [{\citenamefont {Vishnevsky}\ \emph {et~al.}(2013)\citenamefont
  {Vishnevsky}, \citenamefont {Flayac}, \citenamefont {Nalitov}, \citenamefont
  {Solnyshkov}, \citenamefont {Gippius},\ and\ \citenamefont
  {Malpuech}}]{Vishnevsky2013}%
  \BibitemOpen
  \bibfield  {author} {\bibinfo {author} {\bibfnamefont {D.~V.}\ \bibnamefont
  {Vishnevsky}}, \bibinfo {author} {\bibfnamefont {H.}~\bibnamefont {Flayac}},
  \bibinfo {author} {\bibfnamefont {A.~V.}\ \bibnamefont {Nalitov}}, \bibinfo
  {author} {\bibfnamefont {D.~D.}\ \bibnamefont {Solnyshkov}}, \bibinfo
  {author} {\bibfnamefont {N.~A.}\ \bibnamefont {Gippius}}, \ and\ \bibinfo
  {author} {\bibfnamefont {G.}~\bibnamefont {Malpuech}},\ }\href {\doibase
  10.1103/PhysRevLett.110.246404} {\bibfield  {journal} {\bibinfo  {journal}
  {Phys. Rev. Lett.}\ }\textbf {\bibinfo {volume} {110}},\ \bibinfo {pages}
  {246404} (\bibinfo {year} {2013})}\BibitemShut {NoStop}%
\bibitem [{\citenamefont {High}\ \emph {et~al.}(2012)\citenamefont {High},
  \citenamefont {Leonard}, \citenamefont {Hammack}, \citenamefont {Fogler},
  \citenamefont {Butov}, \citenamefont {Kavokin}, \citenamefont {Campman},\
  and\ \citenamefont {Gossard}}]{High2012}%
  \BibitemOpen
  \bibfield  {author} {\bibinfo {author} {\bibfnamefont {A.~A.}\ \bibnamefont
  {High}}, \bibinfo {author} {\bibfnamefont {J.~R.}\ \bibnamefont {Leonard}},
  \bibinfo {author} {\bibfnamefont {A.~T.}\ \bibnamefont {Hammack}}, \bibinfo
  {author} {\bibfnamefont {M.~M.}\ \bibnamefont {Fogler}}, \bibinfo {author}
  {\bibfnamefont {L.~V.}\ \bibnamefont {Butov}}, \bibinfo {author}
  {\bibfnamefont {A.~V.}\ \bibnamefont {Kavokin}}, \bibinfo {author}
  {\bibfnamefont {K.~L.}\ \bibnamefont {Campman}}, \ and\ \bibinfo {author}
  {\bibfnamefont {A.~C.}\ \bibnamefont {Gossard}},\ }\href@noop {} {\bibfield
  {journal} {\bibinfo  {journal} {Nature}\ }\textbf {\bibinfo {volume} {483}},\
  \bibinfo {pages} {584} (\bibinfo {year} {2012})}\BibitemShut {NoStop}%
\bibitem [{sup()}]{suppl}%
  \BibitemOpen
  \href@noop {} {}\bibinfo {note} {See Supplemental Material at [URL will be
  inserted by publisher].}\BibitemShut {Stop}%
\end{thebibliography}%

\end{document}